\documentclass[prl,twocolumn, superscriptaddress]{revtex4}
\usepackage{graphicx,amsmath,amssymb,amsxtra}
\def\bd{\begin{displaymath}}
\def\be{\begin{equation}}
\def\ed{\end{displaymath}}
\def\ee{\end{equation}}
\def\bsub{\begin{subequations}}
\def\esub{\end{subequations}}

\newcommand{\Eq}[1]{Eq.~(\ref{#1})}
\newcommand{\Fig}[1]{Fig.~\ref{#1}}

\makeatletter
\newcommand*\dashline{\rotatebox[origin=c]{90}{$\dabar@\dabar@\dabar@$}}
\makeatother

\newcommand{\ket}[1]{\ensuremath{|#1\rangle}}
\newcommand{\bra}[1]{{\langle #1|}}

\begin{document}

\title{Correlation Functions in SU(2)-Invariant Resonating-Valence-Bonds Spin Liquids on Nonbipartite Lattices}

\author{Julia Wildeboer}

\affiliation{
Department of Physics and Center for Materials Innovation, 
Washington University, St. Louis, MO 63130, USA}

\author{Alexander Seidel}

\affiliation{
Department of Physics and Center for Materials Innovation,
Washington University, St. Louis, MO 63130, USA}

\begin{abstract}
We introduce a  Monte Carlo scheme based on sampling of Pfaffians to 
investigate Anderson's resonating-valence-bond (RVB) spin liquid wave function on the 
kagome and the triangular lattice. This eliminates a sign problem that 
prevents utilization of the valence bond basis in Monte Carlo studies for non-bipartite lattices.
Studying lattice sizes of up to 600 sites, we calculate singlet-singlet and spin-spin
correlations, and demonstrate how the lattice symmetry is restored within
each topological sector as the system size is increased. Our findings are consistent
with the expectation that the nearest neighbor RVB states describe a topological
spin liquid on these non-bipartite lattices. 
\end{abstract}


\maketitle

{\em Introduction.} It has been almost four decades since Anderson proposed\cite{anderson1} the 
quantum spin liquid state. Its undiminished appeal stems from a variety of 
applications from high temperature superconductivity\cite{anderson2} to quantum computing \cite{kitaev,freedman0}. The nature 
of the short ranged variant of Anderson's ``resonating valence bond'' 
(RVB) spin liquid as a topological phase became understood through 
a series of papers\cite{thoulessPRB87,read_chak, kivelsonPRB89}. In particular, the invention of quantum 
dimer models\cite{kivelsonPRB89} as an approximation to spin models finally lead 
to a lattice model exhibiting a topological RVB liquid phase\cite{MS, misguich}.
This however, did not immediately address the (original) question whether 
this exotic phase could be stabilized within the phase diagram 
of $SU(2)$-invariant local spin-$1/2$ Hamiltonians. 
This was subsequently established for highly decorated lattices\cite{MRS} and certain bipartite lattices
\cite{fujimoto}, by finding a parent Hamiltonian for the simplest, i.e., nearest neighbor 
version of the prototypical RVB spin liquid wave function on such lattices.
Work on quantum dimer models\cite{Sachdev89, Leung, MSFradkin, MS}, however, strongly suggests that 
nearest neighbor RVB states should be critical on bipartite lattices, as demonstrated recently\cite{alet,sandvik}. 
They should describe a $\mathbb{Z}_2$-spin liquid with exponentially decaying correlations 
only in the nonbipartite case. While rigorously proven in 
the quantum dimer case, it is highly non-trivial establish 
this statement for the spin-$1/2$ RVB wave functions, 
due to orthogonality issues (cf, e.g., \cite{wildeboer_seidel}). 
In the nonbipartite case, 
the nature of the correlation functions of the local spin 
and valence bond operator has not yet been studied systematically.
This is largely due to a sign problem that will 
be addressed in this work. We finally mention that 
for the kagome case, the short-ranged RVB state studied here has 
been proven to be the ground state of a local parent 
Hamiltonian\cite{seidelkagome} (cf. also \cite{cano, yao}).
The present work will provide essential evidence from correlations demonstrating
that the kagome lattice  RVB ground state of the Hamiltonian 
given in \cite{seidelkagome} is a topological ($\mathbb{Z}_2$) spin liquid.
\begin{figure}[t]
\centering
\includegraphics[width=8cm]{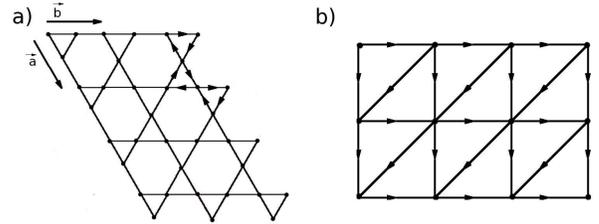}
\caption{ 
a) Shape of the kagome lattice used in the calculations. The lattice consists of $m$ unit cells in the
$\mathbf{a}$ direction and $n$ unit cells in the $\mathbf{b}$ direction, for a total of $3mn$ sites.
Periodic boundary conditions may or may not be introduced with periods $m\mathbf{a}$
and $n\mathbf{b}$. 
b) The orientation used in the sign convention for the triangular lattice.
\label{orient_triangular}
 }
\end{figure}

{\em Method.}
The standard method for calculating correlations of these wave functions on bipartite
lattices is based on the observation that a general correlator between two 
local operators ${\cal O}_1$ and $ {\cal O}_2$ takes on the form
\begin{eqnarray}\label{corr}
\cfrac{\langle RVB|{\cal O}_1 {\cal O}_2|RVB\rangle}{\langle RVB|RVB\rangle}=\cfrac{\sum_{D,D'}\langle D|{\cal O}_1 {\cal O}_2|D'\rangle}{\sum_{D,D'}\langle D|D'\rangle}\,.
\end{eqnarray}
Here $D$ and $D'$ represent dimerizations of the lattice, the sums run over all possible dimerizations,
\ket{D} is a nearest-neighbor valence bond (NNVB) state associated with a given dimerization and a link orientation
of the lattice (defined below), and \ket{RVB} is the RVB state, $\ket{RVB}=\sum_D \ket{D}$.
Since every pair of dimer configurations $D, D'$ corresponds to a configuration of non-intersecting
close packed loops on the lattice, Sutherland pointed out 
\cite{sutherland}
that the evaluation of such correlation functions may be reduced to the study of a classical loop gas
model. This is so {\em provided that} the overlaps $\langle D|D'\rangle$
are strictly non-negative. Otherwise, the evaluation of these correlators through Monte Carlo methods suffers from
a sign problem. Indeed, it is not difficult to show that, e.g., for the kagome lattice, for any sign convention
for the states $|D\rangle$
some of the overlaps $\langle D|D'\rangle$ are always negative.
\begin{figure}[t]
\centering
\includegraphics[width=7cm]{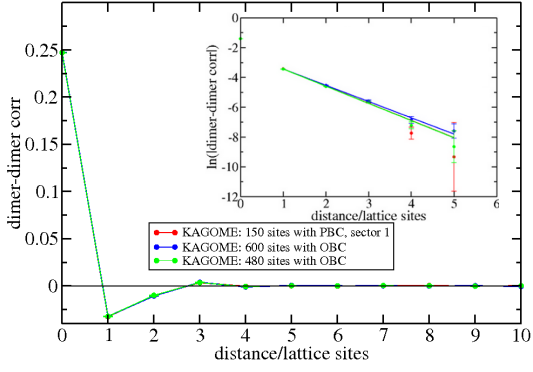}
\caption{ 
The dimer-dimer correlation function shown for kagome lattices with PBCs and OBCs. 
Insensitivity to system size and short correlation length are evident.
The PBC case has been calculated within a fixed topological sector. The inset shows a logarithmic plot including a linear fit,
yielding a correlation length
 of  $1.12(3)$.  
\label{dimercorr_kagome_obcpbc}
 }
\end{figure}
\begin{figure}[b]
\centering
\includegraphics[width=7cm]{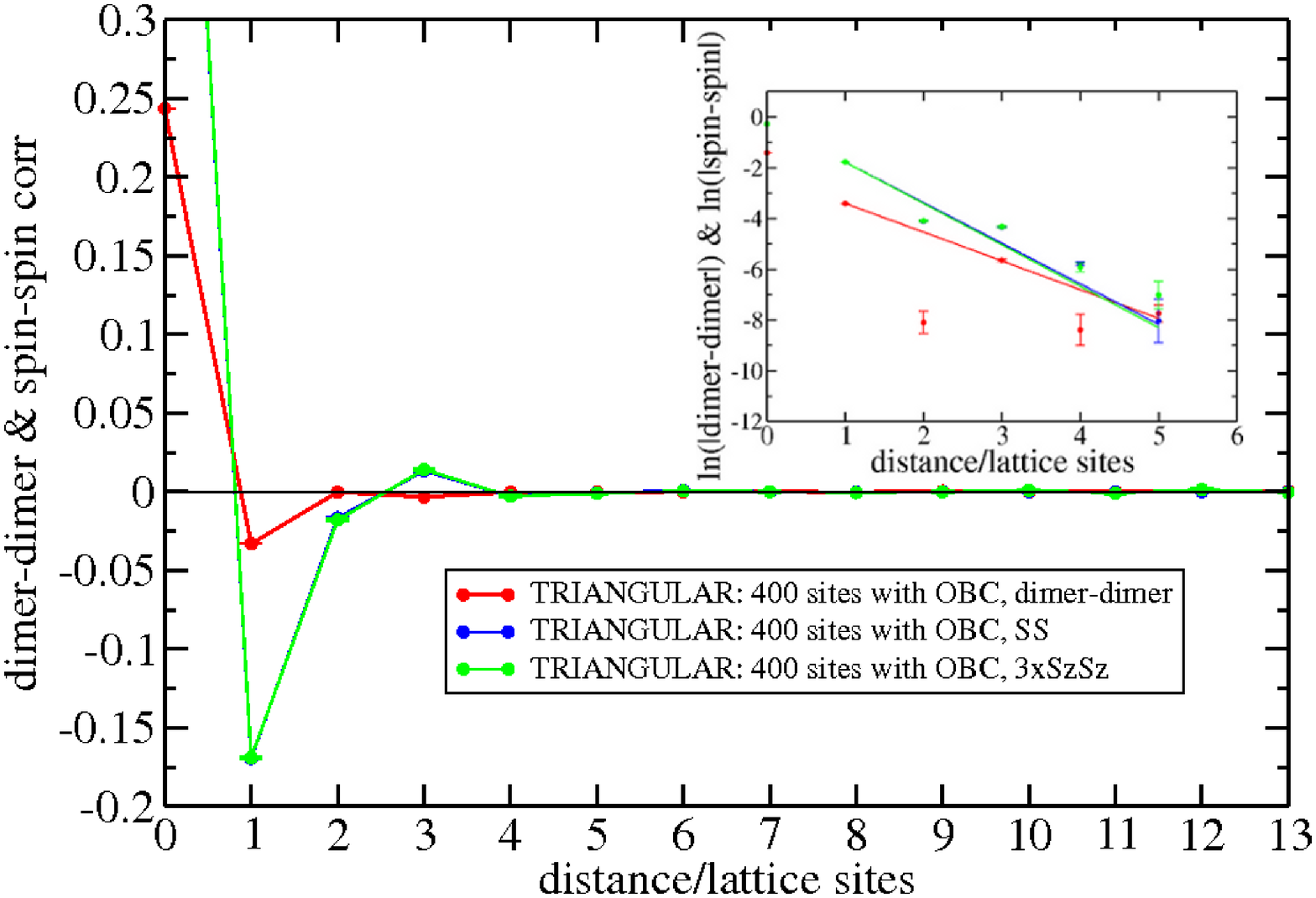}
\caption{ 
The dimer-dimer and  spin-spin correlation functions for a 400 sites triangular lattice with OBCs. 
The inset shows logarithmic plot with fits, giving a correlation length of $1.15(2)$ for the 
dimer-dimer decay. The spin-spin correlations display stronger even/odd effects at short distance.
Fitting only odd distances
in the spin-spin case gives a correlation length of $1.61(2)$. 
\label{dimercorrtriangularobc}
 }
\end{figure}
\begin{figure}[t]
\centering
\includegraphics[width=7cm]{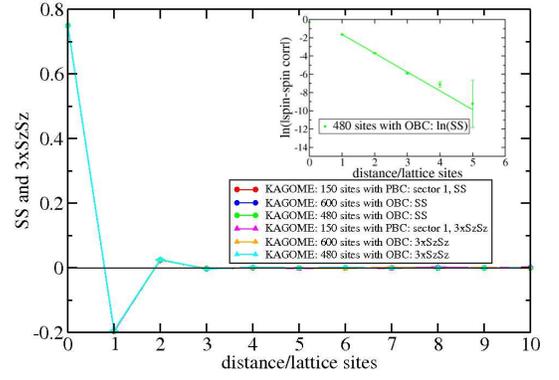}
\caption{ 
The spin-spin correlation functions $\langle \vec{S}_{i}\vec{S}_{i+\kappa} \rangle$ and $\langle S^{z}_i S^{z}_{i+\kappa}\rangle$ 
for different kagome lattices (PBC and OBC). Again, the topological sector was fixed in the PBC case.
The inset shows a logarithmic plot with linear fit yielding a correlation length of $2.08(2)$. 
\label{corrSSSzSzkagomepbcobc}
 }
\end{figure}
\begin{figure}[t]
\centering
\includegraphics[width=7.5cm]{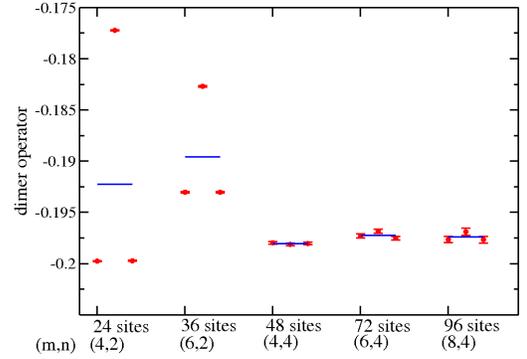}
\caption{ \label{symmetry}
The expectation value of the dimer operator for links of 
the three possible directions and various lattice sizes. The average for one system size is shown
as horizontal bar. A topological sector has been fixed.
The discrepancy between inequivalent links rapidly decreases with system size, restoring the lattice symmetry. 
 }
\end{figure}
It is clear that such a sign problem would never arise were we to work with an orthogonal basis. In this
case only strictly positive diagonal terms appear in the denominator of the expression replacing \Eq{corr}.
An obvious candidate for such a basis is the ``Ising''-basis where local spins have definite $z$ projection,
$|RVB\rangle=\sum_{I} a_{I} |I\rangle$,
and $I$ runs over all possible Ising spin configurations.
Two primary questions need to be addressed to determine whether the Ising-representation 
lends itself to Monte Carlo evaluation of correlations. The first is the obvious question 
whether for the wave function \ket{RVB}, the coefficients $a_I$ in the above representation
can be efficiently calculated. The second question relates to the observation that for the short-ranged RVB state \ket{RVB}, 
it turns out that only a small fraction of configurations $I$ will lead to non-zero $a_I$. 
One may, however, ask  if once
an $I$ with non-zero $a_I$ is found, a sufficiently local update of $I$ has a high chance of leading 
to a new $I'$ with $a_{I'}\neq 0$. 
To proceed, we first need to express the wave function \ket{RVB} in the Ising basis.
We observe that in the Ising basis, the wave function \ket{RVB} can be naturally written as a ``Haffnian''\cite{supp}:
\begin{eqnarray}\label{aI}
a_{I} = \mbox{Haff}[M_{ij}(I)] \equiv\frac{1}{2^{N/2}(\frac{N}{2}!)}\times\nonumber\\
\sum_{\lambda \in S_N} M_{\lambda_1 \lambda_2}(I) M_{\lambda_3 \lambda_4}(I) \times \cdots \times 
M_{\lambda_{N-1} \lambda_N}(I) \,.
\end{eqnarray}
Here, $M$ is a symmetric matrix whose indices run over the $N$ lattice sites and which depends on the Ising 
configuration via $M_{ij}(I)= \Theta_{ij}( \delta_{\sigma_{i},\uparrow} \delta_{\sigma_{j},\downarrow} - 
\delta_{\sigma_{i},\downarrow} \delta_{\sigma_{j},\uparrow})$, where the $\sigma_{i}$
describe the Ising configuration $I$. $\lambda$ runs over all permutations
of the $N$ sites, and $\Theta_{ij}$ is the matrix describing the chosen orientation of the lattice.
An orientation refers to a relation defined between any two nearest-neighbor lattice sites $i$, $j$,
according to which either $i<j$ or $i>j$ holds. Then $\Theta_{ij}=0$ if $i$, $j$ are not nearest neighbors,
$\Theta_{ij}=1$ for $i>j$, and $\Theta_{ij}=-1$ for $i<j$. Here we consider the orientation
chosen for the kagome lattice indicated by the arrows in \Fig{orient_triangular}a.
The formal definition of the Haffnian is related to that of the 
Pfaffian 
through omission of the sign factor $(-1)^\lambda$.
While the Pfaffian and the determinant can be evaluated in polynomial time, it is not known
how to do this for the other two cases. It would thus be desirable to rewrite \Eq{aI} through a Pfaffian.
Luckily, this is the same problem that Kasteleyn solved long ago\cite{kasteleyn}, which has been
a standard tool in the study of classical and quantum dimer models. In the present context,
it does not seem to have enjoyed much attention. Kasteleyn evaluated the partition function of the
classical dimer gas, which is exactly \Eq{aI} with $M_{ij}$ replaced by $|\Theta_{ij}|$.
He found that this problem may be written as $\mbox{Pfaff}[|\Theta_{ij}|\Theta^K_{ij}]$
where $\Theta^K$ is a matrix similar to $\Theta$, but describing a different, so-called ``Kasteleyn'' 
orientation of the lattice. For planar lattice graphs, such an orientation may generally be found.
A Kasteleyn orientation for the kagome lattice is given in \cite{wangwu}.
The same method works in \Eq{aI} \cite{supp}. We thus have
$a_I= \mbox{Pfaff}[M_{ij}(I)\Theta^K_{ij}]$.
We are now in a position to cast the problem of evaluating the correlation functions \eqref{corr}
as a {\it classical} statistical mechanics problem.  We have:
\begin{eqnarray}
  \label{corr2}
\frac{\langle RVB| {\cal O}_i {\cal O}_j \ket{RVB}}{\bra{RVB}RVB \rangle}  
&=& 
\frac{\sum_I \sum_{I'} a_{I} a_{I'} \bra{I'} {\cal O}_i{\cal O}_j \ket{I}}{\sum_{I} |a_{I}|^2} \nonumber \\ 
&=& 
\frac{\sum_{I} |a_{I}|^2 \sum_{I'} \frac{a_{I'}}{a_{I}} \bra{I'} {\cal O}_i {\cal O}_j \ket{I} }{\sum_{I} |a_{I}|^2}\,.  \nonumber \\
& &
\end{eqnarray}
This may now be interpreted as the classical expectation value 
of a quantity $f$:
$\langle f \rangle = {\sum_{I} f_{I} e^{-E_I}}/{\sum_{I} e^{-E_I}}$.
Here, $e^{-E_I} = |a_I|^2$ and the value $f_I$ of the quantity $f$ in the Ising configuration $I$ is given by 
$f_{I} = \sum_{I'} \bra{I'}{\cal O}_i {\cal O}_j \ket{I} \frac{a_{I'}}{a_I}$. 

We have now demonstrated that the evaluation of correlation functions can be cast
in terms of a partition function, whose weights are positive and can be evaluated in polynomial time
(the structure of our Pfaffian in fact allows reduction to the determinant of an $N/2\times N/2$ matrix).
Returning to our earlier caveat, we moreover found 
that once we have an initial Ising configuration $I$ with $a_I\neq 0$, performing updates\cite{supp}
by exchanging
neighboring spins 
has a high chance of leading to a new configuration $I'$ with $a_{I'}\neq 0$,
The basic requirements for Monte Carlo evaluation
are thus met.

{\em Results.}
Simulations are now performed for differnt lattices sizes. For the kagome
lattice, we have chosen $(m,n)$ as defined in \Fig{orient_triangular} to be (10,5) for periodic boundary conditions (PBCs) and
to be 
(20,8) and  (20,10) for open boundary conditions (OBCs), resulting in a total number of
$N=150$, $480$, and $600$ sites, respectively, 
(and in lattices with roughly unit {\em perpendicular}
aspect ratio).
 For the triangular lattice,
we show data belonging to a $20\times 20$ ``square'' with diagonals (see
\Fig{orient_triangular}) giving a lattice of 400 sites.
In one Monte Carlo sweep through the lattice, 
we attempt to do a
number of 
N exchanges of two neighboring spins. 
All expectation values were calculated by
making about 1,500,000 measurements on the configurations produced by the
Monte Carlo process, allowing the system to equilibrate for 8000 sweeps.
Autocorrelation times are generally quite low, on the order of 1.

\Fig{dimercorr_kagome_obcpbc} presents the connected correlation function of the ``dimer'' or valence bond operator $\vec{S}_i\cdot \vec{S}_{i+x}$,
where $i$ and $i+x$ are nearest neighbors, for different lattice sizes and boundary conditions. 
It is evident that there is a finite and very short correlation length.
From the inset it is clear that  the absolute values of the correlation functions follow
a simple exponential law already at short distance, from which we obtain a correlation length of
$\xi=1.12(3)$.
Moreover, the plot for 600 sites and OBCs coincides very well with that for 150 sites and
PBCs. We note that for the case of PBCs, the method used to treat the classical 
dimer case\cite{wangwu} can again be adapted to the present situation, and yields an expression of the amplitude
$a_I$ as a superposition of four Pfaffians. Different such superpositions can be used to project onto different
topological sectors of the toroidal system. While only one topological sector is shown,
we have also convinced ourselves that results for different topological sectors agree within error bars.
The fact that the dimer-dimer correlations are apparently insensitive
to both lattice size {\em and} boundary conditions, already for a relatively small size of 150 sites, is consistent
with the hypothesis of a gapped state. 
We note moreover that the decay is very
reminiscent of the quantum dimer model case, where dimer-dimer correlations have been shown to decay super-exponentially,
with correlations being exactly zero beyond distance 2\cite{Oshi}. While 
this is clearly not the case for the RVB state,  a very short correlation length
of order 1 still mimics this behavior fairly closely.
The qualitative agreement between the quantum dimer model
and the RVB state is thus quite striking. 

\Fig{dimercorrtriangularobc} shows the dimer-dimer correlations
for a 400 site triangular lattice, displaying similarly short ranged correlations.
Subdominant corrections to the dominant exponential decay are clearly somewhat more
important than for the kagome, as one would generically expect; however a correlation
length close to 1.6 is still clearly visible in the inset. All linear fits are obtained from a weighted least 
square regression, where the weights have been chosen as the inverse squares of the error bars.
Note that although the value at distance zero has not been included into any fit, even this 
shortest distance data point
 tends to follow the exponential trend very well. 
 We point out that a sign
convention for the triangular lattice exists which eliminates the sign problem of \Eq{corr}
\cite{MRS_aip}. Here, however, 
we have chosen a different convention (\Fig{orient_triangular}b), for which this
problem persists.

We also computed spin-spin correlation functions $\langle \vec S_i\cdot \vec S_j\rangle$.
Results are shown for the kagome in \Fig{corrSSSzSzkagomepbcobc} and for the triangular lattice in
\Fig{dimercorrtriangularobc}.
Spin-spin correlations decay exponentially even in the critical
square-lattice case\cite{liang}, and by theoretical prejudice should decay
exponentially for all short ranged RVB states. 
Moreover, even on the kagome, DMRG work 
has predicted a spin liquid phase with gapped spin but gapless
singlet excitations\cite{sheng}. This might render the singlet sector 
more crucial in the present context.
Nonetheless, direct demonstration of the exponential decay of spin-spin corelations
is not straightforward, especially in the presence of the sign problem discussed initially.
Again, the short-ranged nature of the correlations is apparent in both cases.
As a consistency check, both $\langle \vec S_i \cdot \vec S_j\rangle$ and 3$\langle  S^z_i  S^z_j\rangle$ are shown, which must agree by $SU(2)$ symmetry. This symmetry is, however, not manifest in the Ising-basis
we are working with. \footnote{
We note that $S^z  S^z$ correlations for the kagome were also calculated very recently 
using a PEPS representation \cite{schuch}. However, we find a direct comparison of our results not 
straightforward, due to the blocking procedure carried out in \cite{schuch}. }

Up to now we have demonstrated that connected correlations for the RVB states
on the kagome and triangular lattice are short-ranged. This does, however, by itself
not guarantee the liquid property of these states. 
In particular, the four degenerate RVB ground states on the torus transform nontrivially
under the space group of the lattice, 
and to demonstrate the liquid property and rule out the possibility of a valence bond solid \cite{shastry}, 
it is essential to show that the full lattice
symmetry is restored in the thermodynamic limit, for each individual ground state (within each topological sector). 
 We restrict ourselves to the kagome lattice here. In the following, we will refer to lattice links
 as ``symmetry inequivalent'' if they are not related by a symmetry of the
 {\em wave function}  (even though they may be related by a symmetry of the lattice).
 For lattices of the shape shown in (\Fig{orient_triangular}a), with $m$, $n$ both even,
any three links along different directions will always exhaust all possible classes of 
inequivalent links. In \Fig{symmetry}, we plot the expectation values of the dimer operator
for three such links, evaluated in one topological sector, for various ``even/even'' lattices.
One observes that the discrepancy between inequivalent links rapidly decreases,
by a factor of {\em at least} 60  between 24 sites and 48 sites, taking into account error bars.
(The consistency between symmetry equivalent links suggests that the error is much smaller than
shown, and the factor is really on the order of 100).
 For larger lattice size, the calculation becomes increasingly demanding, since, presumably,
increasingly smaller error bars are needed to resolve the discrepancy in expectation values,
while even maintaining the size of the error bars is more costly (\Fig{symmetry}).
It is worth noting, though,
that the average of the three expectation values for 72 and 96 sites appears to have converged, and 
we are thus approaching the thermodynamic limit. 
In all, these findings are highly consistent with the general expectation
that the RVB-states describe a topological spin liquid.
 
 {\em Conclusion.}
 In this work, we have studied correlation functions of nearest neighbor resonating valence bond
 wave functions on both the kagome and the triangular lattice, with up to 600 lattice sites.
 A sign problem of earlier methods has been circumvented by using a Pfaffian representation
 of the wave function in the Ising basis. This allows for evaluation of correlators for both OBCs
 and PBCs, and, in the latter case, restriction to a single topological sector.
 This allowed us to present strong evidence that not only correlations decay exponentially as expected, but also
 that no broken lattice symmetry remains in the thermodynamic limit for the kagome lattice.
 For the kagome, this greatly adds to the amassed evidence
  that local $SU(2)$ invariant Hamiltonians
 stabilizing a topological spin liquid state are possible\cite{seidelkagome,schuch}. 
 Further possible applications of our method include the investigation of short-ranged RVB wave functions 
 on other nonbipartite lattices. In particular, certain next nearest neighbor links may be 
 introduced in standard lattice geometries such as the square lattice\cite{moessner_sandvik},
 as long as the planarity of the lattice is maintained. This makes it natural to introduce different weights
 for different types of valence bonds. 
 Furthermore, 
 our method allows for the introduction of any number of mobile (delocalized) holes 
  and thus the study of monomer correlations and the related confinement/deconfinement issue.
  We are hopeful that these prospects will stimulate future work.
 \begin{acknowledgments}
 We would like to thank M. Ogilvie for insightful discussions. This work has been supported by the 
 National Science Foundation under NSF Grant No. DMR-0907793. Our MC codes 
 are partially based upon the ALPS libraries\cite{alps1,alps2}.
 \end{acknowledgments}

 {\em Note added.} 
 After the completion of this work, Ref.[34] appeared using an approach
that has been outlined in Ref.[35].


\medskip 
{ \bf Supplemental material: Haffnian and Pfaffian representation of RVB state, and ergodicity}\\

We begin with the Haffnian representation of the RVB state 
$|RVB\rangle=\sum_{I} a_{I} |I\rangle\;.$
We reproduce Eq. (2) of the paper as
\begin{eqnarray}\label{Haff}
a_{I} = \mbox{Haff}[M_{ij}(I)] \equiv\frac{1}{2^{N/2}(\frac{N}{2}!)}\times\nonumber\\
\sum_{\lambda \in S_N} M_{\lambda_1 \lambda_2}(I) M_{\lambda_3 \lambda_4}(I) \times \cdots \times 
M_{\lambda_{N-1} \lambda_N}(I) \,.
\end{eqnarray}
Independent of $I$, $|M_{ij}(I)|$ is just the adjacency matrix of the lattice, the only contributions come from
permutations $\lambda$ such that $(\lambda_{2n-1}\lambda_{2n})$ is a nearest neighbor pair, $n$ being an integer running from
$1$ to half the number of lattice sites. Therefore, a contributing permutation $\lambda$ corresponds to a dimerization $D$ of the lattice 
into nearest neighbor pairs. The permutations corresponding to the same dimerization in this way are all related by permuting the members of the individual pairs and by permuting pairs among each other.
This over-counting is compensated by the combinatorial prefactor.
The sum in \Eq{Haff} is thus essentially a sum over dimer configurations $D$.
Since as function of $\sigma_{\lambda_{2n-1}}$ and  $\sigma_{\lambda_{2n}}$, 
$M_{\lambda_{2n-1} \lambda_{2n}}(I)$ is just the wave function of a singlet
occupying the link $(\lambda_{2n-1}\lambda_{2n})$, endowed with the correct sign,
the contribution of a  given $\lambda$  to \Eq{Haff} is just the spin wave function associated
with the dimer covering $D$ in the RVB state. 

On the other hand, the Pfaffian representation can be written as

\begin{eqnarray}\label{Pfaff}
a_{I} = \mbox{Pfaff}[\tilde M_{ij}(I)] \equiv\frac{1}{2^{N/2}(\frac{N}{2}!)}\times\nonumber\\
\sum_{\lambda \in S_N}  (-1)^\lambda \Theta^K_{\lambda_1 \lambda_2}(I) \Theta^K_{\lambda_3 \lambda_4}(I) \times \cdots \times 
\Theta^K_{\lambda_{N-1} \lambda_N}(I)  \nonumber\\
M_{\lambda_1 \lambda_2}(I) M_{\lambda_3 \lambda_4}(I) \times \cdots \times 
M_{\lambda_{N-1} \lambda_N}(I) \,.
\end{eqnarray}

For reasons of self-containedness, we reproduce here the well-known arguments\cite{kasteleyn1} that Eqs. \eqref{Haff} and \eqref{Pfaff} are interchangeable.
This also serves to make it manifest that these arguments carry over from the original classical dimer problem to the present situation without any
difficulty. We first want to see that  Eqs. \eqref{Haff}, \eqref{Pfaff} are identical within each topological sector of dimer coverings, possibly up to
an overall negative sign. To this end, we use the fact that all coverings within a topological sector are mutually related by a sequence of ``resonance moves'', where during a single resonance move, a set of dimers is shifted along a path of links forming a contractible (topologically trivial)
loop. This interchanges occupied and unoccupied links along the loop. It is therefore sufficient to show that the additional term
of \Eq{Pfaff} compared to \Eq{Haff} is constant under a resonance move. Without any loss of generality, we can consider permutations
$\lambda$, $\lambda'$ corresponding to the dimer covering before and after the resonance move, respectively, that differ by a single 
cyclic permutation of the values of  $\lambda_1\dotsc \lambda_{2k}$. Then $(-1)^\lambda=-(-1)^{\lambda'}$.
Moreover, the defining property of the Kasteleyn orientation is that the number of clockwise oriented links along any
contractible loop of even length, which encloses an even number of lattice sites, is odd. (To ensure this, it  is sufficient that the number of clockwise oriented links is odd around 
any elementary plaquette of the planar lattice graph,  for both even and odd plaquette perimeter).
This then implies that the same cyclic permutation of indices described above also introduces an additional minus sign 
into the product $\Theta^K_{\lambda_1,\lambda_2}\cdot\dotsc\cdot \Theta^K_{\lambda_{2k-1},\lambda_{2k}}$.
This shows that \eqref{Haff} and \eqref{Pfaff} are identical within a topological sector up to a sign.
For open geometry, there is just one topological sector (up to ``non-resonable" configurations in the triangular lattice case\cite{MS1}
that are of negligible weight in the RVB state), and we are done.
For periodic boundary conditions (torus geometry), Eqs. \eqref{Haff} and \eqref{Pfaff} do in general differ by a sign 
that is different for 
different topological sectors. However, this relative sign can be changed by further modifying $\Theta^K$ along topologically
non-trivial loops\cite{wangwu1}, making it possible to render all signs identical, or,
 as we did, to project onto a single topological sector by taking superpositions for
different such modified $\Theta^K$.

These considerations also set us up to observe an ergodicity property that the nearest neighbor 
spin exchange moves employed in our Monte Carlo method have.
Stated precisely, these moves are ergodic within the subset of Ising basis states
that have non-zero weight for at least one nearest neighbor valence bond state (NNVB)
within a given topological sector. First of all, it is obvious that nearest neighbor
exchanges connect all Ising configurations belonging to a given NNVB state: Such configurations
are linked by exchanges on links belonging to singlet pairs in the NNVB state.
Moreover, for every dimer loop in the dimer covering associated with the NNVB state,
there are Ising configurations contributing to this state that are Neel ordered along this loop.
Such Ising configurations, however, have non-zero weight not only in this particular NNVB state,
but also in the NNVB state whose dimer configuration differs from the original
one by a single resonance move along this particular dimer loop. Then, by induction 
over the minimum number of resonance moves needed to connect two dimer coverings
in the same topological sector,
all Ising basis states having non-zero weight in the NNVB states associated with these
coverings can be connected by nearest neighbor spin exchanges.
 

\end{document}